# Ferromagnetism in Cr-doped topological insulator TlSbTe$_2$

Zhiwei Wang, Kouji Segawa[†], Satoshi Sasaki[§], A. A. Taskin, and Yoichi Ando[*]

*Institute of Scientific and Industrial Research, Osaka University, Ibaraki, Osaka 567-0047, Japan*

We have synthesized a new ferromagnetic topological insulator by doping Cr to the ternary topological-insulator material TlSbTe$_2$. Single crystals of Tl$_{1-x}$Cr$_x$SbTe$_2$ were grown by a melting method and it was found that Cr can be incorporated into the TlSbTe$_2$ matrix only within the solubility limit of about 1%. The Curie temperature $\theta_c$ was found to increase with the Cr content but remained relatively low, with the maximum value of about 4 K. The easy axis was identified to be the *c*-axis and the saturation moment was 2.8 $\mu_B$ (Bohr magneton) at 1.8 K. The in-plane resistivity of all the samples studied showed metallic behavior with *p*-type carriers. Shubnikov-de Hass (SdH) oscillations were observed in samples with the Cr-doping level of up to 0.76%. We also tried to induce ferromagnetism in TlBiTe$_2$ by doping Cr, but no ferromagnetism was observed in Cr-doped TlBiTe$_2$ crystals within the solubility limit of Cr which turned out to be also about 1%.

[†]Present address: Department of Physics, Kyoto Sangyo University, Motoyama, Kamigamo, Kita-ku, Kyoto 603-8555, Japan

[§]Present address: School of Physics and Astronomy, University of Leeds, Leeds LS2 9JT, United Kingdom

[*]Electronic mail: y_ando@sanken.osaka-u.ac.jp

Three-dimensional (3D) topological insulators (TIs) are a class of materials characterized by a nontrivial $Z_2$ topology of the bulk wave function, where an insulating bulk hosts a linearly dispersing surface state protected by the time-reversal symmetry.[1-8] Theoretical studies showed that the thallium-based ternary chalcogenides $TlSbTe_2$, $TlBiSe_2$, and $TlBiTe_2$ are 3D TIs with a single-Dirac-cone surface state at the Γ point.[9-11] Experimentally, $TlBiSe_2$ and $TlBiTe_2$ have been confirmed to be topological insulators;[12-14] in particular, it was found that the surface-state structure of $TlBiSe_2$ is similar to that in $Bi_2Se_3$, making it suitable for studying the Dirac-cone physics in a simple setting with a large bulk band gap or ~0.35 eV.[12] Interestingly, a topological phase transition was found in the $TlBi(Se_xS_{1-x})_2$ solid solution,[15, 16] which provides a platform for realizing the 3D Dirac semimetal. Furthermore, if one could induce ferromagnetism at the topological phase transition point of this solid-solution system by doping a magnetic element, the broken time-reversal symmetry would lead the emergence of a Weyl semimetal.[17] Theoretical studies have also shown that quantum anomalous Hall effect could occur in Tl-based TIs when doped with transition metals (TMs).[18] Therefore, TM doping to Tl-based TIs would be important for the pursuit of novel quantum states of matter.

In the past, TM doping to various TI materials has been tested: Successful observations of ferromagnetism were reported for Mn-doped $Bi_2Te_3$,[19] V- or Cr-doped $Sb_2Te_3$,[20,21] Fe-doped $Bi_2Te_3$,[22] and V- or Cr-doped $(Bi,Sb)_2Te_3$.[23,24] So far, such ferromagnetic TIs have been found only in tetradymite TI materials and ferromagnetism has never been observed in a Tl-based ternary TI. Here we report our explorations of ferromagnetism in $TlSbTe_2$ and $TlBiTe_2$ by TM doping. We found that Cr doping can induce ferromagnetism in $TlSbTe_2$, but not in $TlBiTe_2$.

The single crystals of Cr-doped $TlSbTe_2$ were grown by a melting method using elemental shots of Tl (99.99%), Sb (99.9999%), and Te (99.9999%) as well as powders of Cr (99.9%) as starting materials. Mixtures of those materials with the nominal composition of $Tl_{1-x}Cr_xSbTe_2$ ($x$ = 0.00, 0.007, 0.01, 0.02, 0.03) were prepared with the total weight of 4.0 g and were sealed in evacuated quartz tubes; we also prepared a batch of Te-rich composition, $Tl_{0.98}Cr_{0.02}SbTe_{2.2}$, for comparison. The quartz tubes were heated and kept at 700°C for 48 h with intermittent shaking to ensure homogeneity of the melt, followed by cooling slowly to 450°C. Single crystals with the lateral dimension of up to a few centimeters can be obtained by cleaving along the (00l) plane. We also synthesized $Tl_{1-x}Cr_xBiTe_2$ ($x$ = 0.01, 0.02, 0.03) crystals with the same method (Bi purity was 99.9999%). Note that before the synthesis of our samples, we preformed surface cleaning procedures to remove the oxide layers formed in air on the raw shots of Tl, Sb, and Bi: Tl shots are annealed in hydrogen atmosphere at 230°C for 2 h; Sb and Bi shots are washed with diluted $HNO_3$.

The crystal structure of each sample was checked by powder X-ray diffraction (XRD) using Rigaku Ultima-IV diffractometer with Cu $K\alpha$ emission, which was performed on powders obtained by crushing the crystals. The actual Cr content in the samples was analyzed by inductively-coupled plasma atomic-emission spectroscopy (ICP-AES). Magnetization measurements were carried out using a SQUID magnetometer (Quantum Design MPMS) and a vibrating sample magnetometer (Quantum Design PPMS). The in-plane transport properties were measured in magnetic fields up to 14 T with a standard six-probe method to record the longitudinal resistivity $\rho_{xx}$ and the Hall resistivity $\rho_{yx}$ simultaneously. The single crystal samples for transport measurements were cut into a rectangular shape with a typical size of 2 $\times$ 0.5 $\times$ 0.2 $mm^3$ and electrical contacts were made by using room-temperature-cured silver paste.

Motivated by a theoretical proposal[18] that ferromagnetism should be induced by TM doping in $TlBiX_2$ (X = Te and Se) and that Cr would be the most promising element to induce ferromagnetic order, we started our explorations by growing $Tl_{1-x}Cr_xBiTe_2$. Figure 1(a) shows the powder XRD patterns of the grown $Tl_{1-x}Cr_xBiTe_2$ samples with nominal $x$ values of 0.01 and 0.02. One can see that the $x$ = 0.01 sample is single phase and all the diffraction peaks can be well indexed to the rhombohedral structure of $TlBiTe_2$ with space group R-3m (we use the hexagonal notation).[25] However, peaks from an impurity phase, $Cr_2Te_3$, shows up in the data for $x$ = 0.02 as indicated by asterisks in Fig. 1(a). This phase is known to be ferromagnetic with the Curie temperature of 165 K.[26] In the magnetization data shown in Fig. 1(b), one can see that the $x = 0.02$ sample indeed presents ferromagnetism below 165 K with a clear magnetic hysteresis [inset of Fig. 1(b)]. On the other hand, no clear ferromagnetism was observed down to 1.8 K in the $x = 0.01$ sample which is free from the $Cr_2Te_3$ impurity phase. Therefore, one may conclude that $TlBiTe_2$ has a relatively low solubility limit of about 1% for Cr and that ferromagnetic order is not established above 1.8 K within this solubility limit. For $TlBiTe_2$, we also tried doping of other TM elements, Mn, Fe, and Ni, but none of them were found to induced ferromagnetism.

After obtaining these negative results on $TlBiTe_2$, we switched to work on $TlSbTe_2$. Although the surface state observation has not been successful for $TlSbTe_2$ by angle-resolved photoemission spectroscopy because of its *p*-type nature, there is no reason to doubt its topological nature. Figure 2 shows the powder XRD patterns of $Tl_{1-x}Cr_xSbTe_2$ which were obtained on crushed crystals. For this $Tl_{1-x}Cr_xSbTe_2$ system, we show the actual $x$ values determined by the ICP-AES analysis (see Table 1) except for the nominal $x = 0.03$ sample which was found to contain the $Cr_2Te_3$ impurity phase; all other samples with $x \leq 0.0092$ are single phase with the expected rhombohedral structure of $TlSbTe_2$ (space group R-3m).[25] The samples with the highest actual composition of Cr in the present series, $x = 0.0092$, were obtained from the batches with the nominal $x$ value of 0.02. This suggests that the solubility limit of Cr in $TlSbTe_2$ is about 1%, which is similar to the case of $TlBiTe_2$. We note that the sample indicated as "$x = 0.0092(TR)$" was grown from the Te-rich nominal composition of $Tl_{0.98}Cr_{0.02}SbTe_{2.2}$ and is expected to contain more Te antisite defects compared to other samples. Indeed, the ICP-AES analysis (Table 1) suggests that some of the Tl sites are occupied by Te in this sample; also, as we show later, its hole density was found to be the highest among the present series. The purpose of growing the $x = 0.0092(TR)$ sample was to see the effect of hole density on the Curie temperature in ferromagnetic samples.[23]

The temperature dependences of the magnetization $M$ measured in 0.1 T are shown in Fig. 3(a) for the single-phase samples of $Tl_{1-x}Cr_xSbTe_2$. The rapid increase of the magnetization at low temperature points to a ferromagnetic ordering. The Curie temperature $\theta_C$ can be determined from the Curie-Weiss law

$$M = \frac{C}{T - \theta_C} + M_0$$

by plotting $1/(M-M_0)$ vs $T$ ($C$ is a constant and $M_0$ is the background determined from the high temperature data). For example, the data for the $x = 0.0049$ sample plotted in this way [Fig. 3(b)] can be well fitted by a straight line, whose intercept on the $T$ axis gives $\theta_C$ of 0.8 K; this $\theta_C$ increases to 2.5 and 3.1 K for $x = 0.0076$ and 0.0092, respectively. The relationship between $\theta_C$ and $x$ for those three samples is shown in the inset of Fig. 3 (b), which shows a nearly linear trend. Importantly, the $x = 0.0092$(TR) sample which is expected to have a higher hole density presented the highest $\theta_C$ of 4.1 K, suggesting that $\theta_C$ is determined not only by the density of local moments but also by the density of mobile carriers which would mediate the coupling between local moments. Similar results were reported for Cr- or V-doped $(Bi_{1-x}Sb_x)_2Te_3$.[23,27]

To corroborate the establishment of ferromagnetism in $Tl_{1-x}Cr_xSbTe_2$, we measured $M(B)$ curves; Fig. 4 shows the data for all the single-phase samples at 1.8 K. Clear magnetic hysteresis was observed in all samples except for $x = 0.0049$; note that $\theta_C$ obtained for $x = 0.0049$ was less than 1 K, and hence a hysteresis is not expected for this sample at 1.8 K. The $x = 0.0092$(TR) sample having the highest $\theta_C$ of 4.1 K presents the largest remnant magnetization of ~0.3 $\mu_B$/Cr and the coercive field $B_C$ of 23 mT. This $B_C$ is comparable to that in Mn-doped $Bi_2Te_3$ (35 mT)[19] and in Cr-doped $Sb_2Te_3$ (10 mT),[20] but is much smaller than that in V-doped $Sb_2Te_3$ (1.2 T).[21] The inset of Fig. 4 shows the $M(B)$ curve at 1.8 K for $x = 0.0092$ measured up to 9 T applied parallel to the $c$-axis; the saturated magnetic moment is 2.8 $\mu_B$/Cr, which is a bit smaller than the expected magnetic moment of $Cr^{3+}$ (3.9 $\mu_B$). Note that Cr is antiferromagnetic[28] and its possible clustering cannot explain the observed ferromagnetism.

Figure 5(a) shows the $M(B)$ curves for $x = 0.0092$(TR) measured at 1.8 K with the magnetic field directions of $B // ab$ and $B // c$, from which one can easily see that the magnetic easy axis is the $c$-axis. This easy axis direction is the same as that reported for Mn-doped $Bi_2Te_3$ (ref. 19) and for V- or Cr-doped $Sb_2Te_3$.[20, 21] We also measured the $M(B)$ curves for $x = 0.0092$(TR) at various temperatures as shown in Fig. 5(b); in these measurements, the sample was first heated to 20 K and then cooled to the target temperature in 0 T to guarantee perfect demagnetization. The hysteresis disappears between 4.5 and 6 K, which is consistent with $\theta_C = 4.5$ K determined from the $M(T)$ data.

Now we briefly discuss the transport data of $Tl_{1-x}Cr_xSbTe_2$. The temperature dependences of $\rho_{xx}$ in 0 T and the magnetic-field dependences of $\rho_{yx}$ at 1.8 K are shown in Fig. 6. The absence of a clear anomalous Hall signal in our $\rho_{yx}(B)$ data is probably due to the very small magnetization associated with ferromagnetism (up to ~0.003 $\mu_B$/f.u. at 1.8 K). It is worth noting that the $x = 0.0092$(TR) sample shows the smallest positive slope of $\rho_{yx}(B)$, which means that the hole density is the largest among all the samples. To be concrete, the hole density $p$ estimated from $1/eR_H$ ($R_H$ is the low-field Hall coefficient and $e$ is the elementary charge) is $3.2\times10^{19}$, $1.8\times10^{19}$, $1.2\times10^{19}$, $0.79\times10^{19}$, and $3.9\times10^{19}$ cm$^{-3}$ for $x = 0.0000$, 0.0049, 0.0076, 0.0092, and 0.0092(TR), respectively. Correspondently, the mobilities for these samples are calculated to be 1502, 1509, 2038, 2232, and 1282 cm$^2$/Vs. The decreasing trend in $p$ with increasing $x$ is reasonable, because $Cr^{3+}$ substitution for $Tl^+$ leads to electron doping. The increase in mobility in samples with higher $x$ [except for $x = 0.0092$(TR)] suggests that the electron-electron scattering is dominant over the impurity scattering on $Cr^{3+}$ ions.

We observed clear Shubnikov-de Hass (SdH) oscillations in samples with $x \leq 0.0076$. Figure 7 shows the oscillations in $d\rho_{yx}/dB$, in which the main oscillation frequency $F$ is 128−137 T and does not change much with $x$; the corresponding hole density (assuming a spherical Fermi surface) is $0.8–0.9 \times 10^{19}$ cm$^{-3}$. As is most obvious in the data for $x = 0.0049$, the SdH oscillations present beating, suggesting the existence of more than one Fermi surfaces with similar sizes.

In conclusion, ferromagnetism was observed in $Tl_{1-x}Cr_xSbTe_2$ but not in $Tl_{1-x}Cr_xBiTe_2$ above 1.8 K. The solubility limit of Cr in both $TlSbTe_2$ and $TlBiTe_2$ is found to be about 1% and the $Cr_2Te_3$ impurity phase appears when the Cr content exceeds this solubility limit. The Curie temperature $\theta_c$ in $Tl_{1-x}Cr_xSbTe_2$ increases with both $x$ and the hole density. The highest $\theta_c$ of about 4 K was observed in $x = 0.0092$(TR) sample with the hole concentration of $3.9 \times 10^{19}$ cm$^{-3}$.

This work was supported by JSPS (KAKENHI 25220708 and 25400328), MEXT (Innovative Area "Topological Quantum Phenomena" KAKENHI), and AFOSR (AOARD 124038).

Table 1: Actual compositions of the Cr-doped TlSbTe$_2$ crystals determined from ICP-AES analysis. Since ICP-AES analysis only gives compositional ratios of the constituent elements, the composition values within each sample are determined by setting their sum to be 4.

| nominal composition | Tl | Cr | Sb | Te |
| --- | --- | --- | --- | --- |
| Tl$_{0.993}$Cr$_{0.007}$SbTe$_2$ | 0.9788 | 0.0049 | 1.0453 | 1.9710 |
| Tl$_{0.99}$Cr$_{0.01}$SbTe$_2$ | 0.9865 | 0.0076 | 1.0219 | 1.9840 |
| Tl$_{0.98}$Cr$_{0.02}$SbTe$_2$ | 0.9816 | 0.0092 | 1.0220 | 1.9873 |
| Tl$_{0.98}$Cr$_{0.02}$SbTe$_{2.2}$ | 0.9025 | 0.0092 | 1.0778 | 2.0106 |

Figure captions

Fig. 1. (a) Powder XRD patterns of the Tl$_{1-x}$Cr$_x$BiTe$_2$ samples with nominal $x = 0.01$ and 0.02; asterisks mark the Cr$_2$Te$_3$ impurity phase and the inset shows a magnified comparison between x = 0.01 and 0.02 for the appearance of the Cr$_2$Te$_3$ peaks (the $x = 0.02$ data in the inset are shifted up by 100 for clarity). (b) Temperature dependences of the dc magnetization measured on these samples in 100 mT; inset shows the magnetic-field dependence of the magnetization at 1.8 K.

Fig. 2. Powder XRD patterns of the series of Tl$_{1-x}$Cr$_x$SbTe$_2$ samples grown in this work; asterisks mark the peaks from the Cr$_2$Te$_3$ impurity phase, which were observed only in samples with nominal $x$ values larger than 0.02. The vertical axis in panel (a) is linear, while that in panel (b) is logarithmic.

Fig. 3. (a) Temperature dependences of the dc magnetization measured on the Tl$_{1-x}$Cr$_x$SbTe$_2$ samples in 100 mT; inset magnifies the data at low temperature. (b) Plots of $1/(M-M_0)$ vs $T$, where $M_0$ is the background determined at high temperature; solid lines are linear fits to the data to determine the Curie temperature $\theta_c$. Inset shows $\theta_c$ as a function of actual $x$ and the dashed line is a liner fit to the data.

Fig. 4. $M(B)$ curves at 1.8 K in $B // c$ for Tl$_{1-x}$Cr$_x$SbTe$_2$ with various $x$ values. Inset shows the magnetization of the $x = 0.0092$ sample in magnetic fields up to 9 T.

Fig. 5. (a) $M(B)$ curves observed in the $x = 0.0092$(TR) sample at 1.8 K for $B // ab$ and $B // c$. (b) $M(B)$ curves for $B // c$ in the same sample at various temperatures.

Fig. 6. (a) Temperature dependences of $\rho_{xx}$ in the Tl$_{1-x}$Cr$_x$SbTe$_2$ crystals. (b) Magnetic-field dependences of $\rho_{yx}$ at 1.8 K.

Fig. 7. $d\rho_{yx}/dB$ as a function of $1/B$ for Tl$_{1-x}$Cr$_x$SbTe$_2$ with $x = 0.0000, 0.0049$ and $0.0076$. Dashed lines indicate the positions of peaks or valleys.

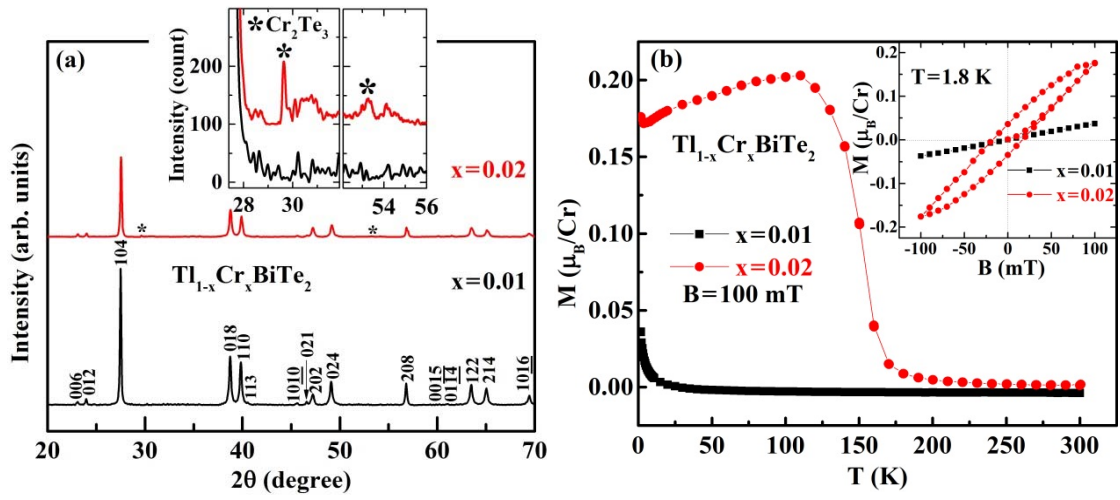

Fig. 1

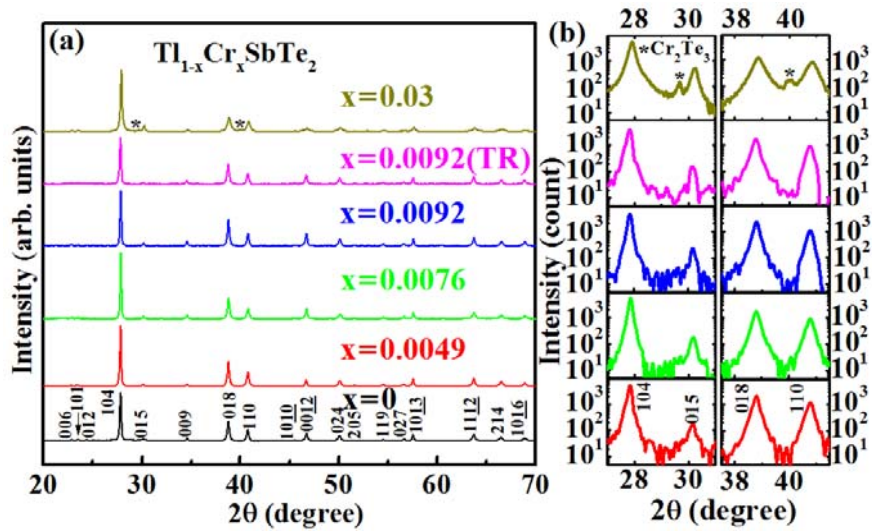

Fig. 2

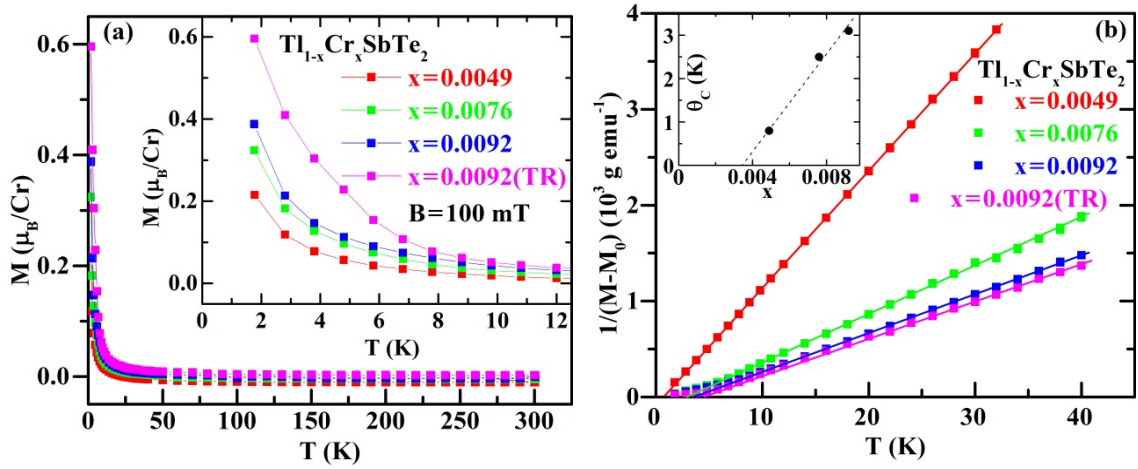

Fig. 3

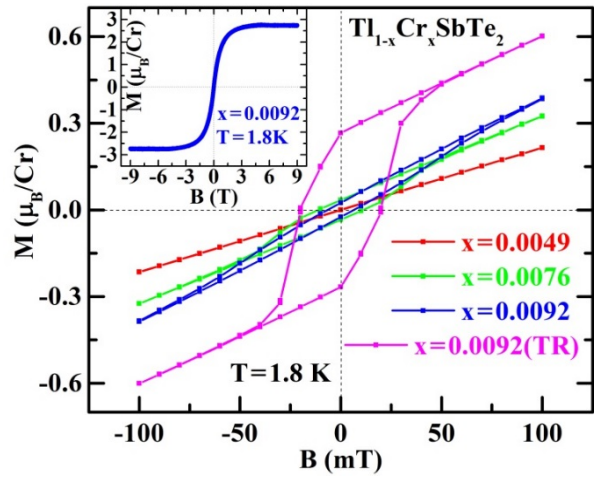

Fig. 4

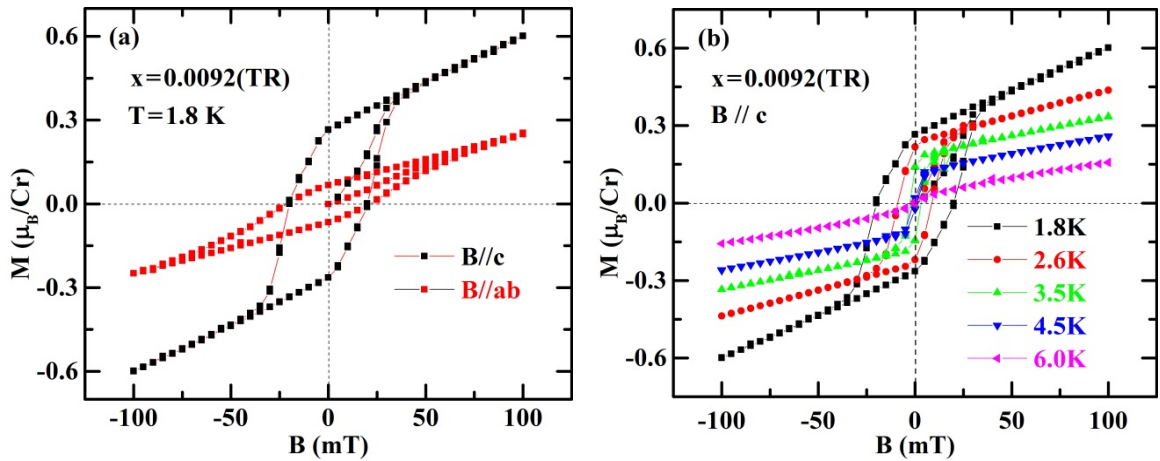

Fig. 5

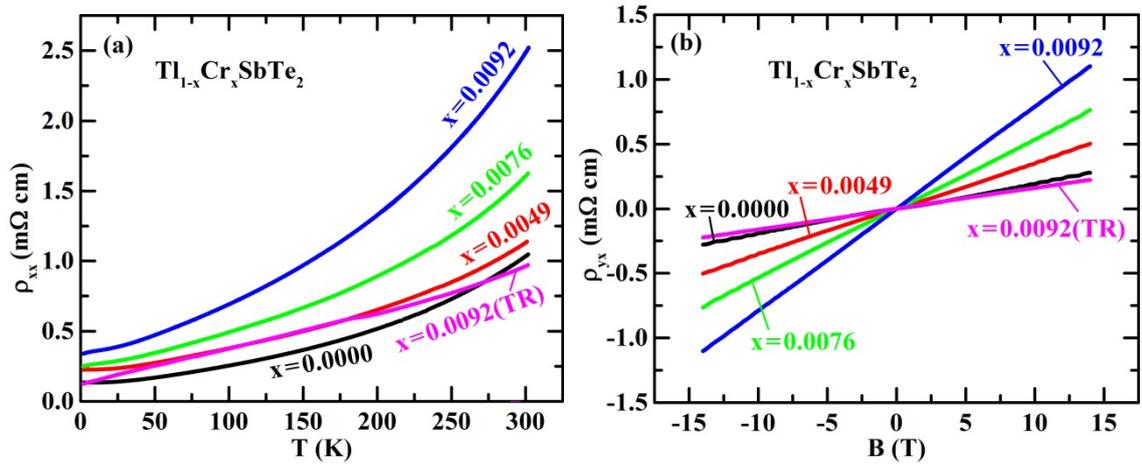

Fig. 6

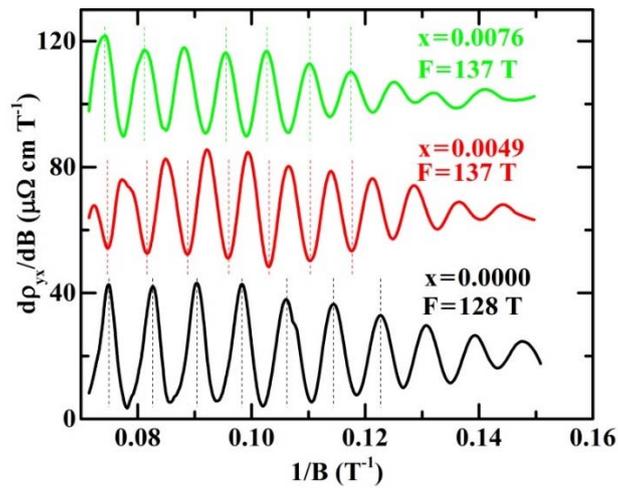

Fig. 7